# Temperature Measurement in Plasmonic Nanoapertures used for Optical Trapping

Quanbo Jiang,[1] Benoît Rogez,[1] Jean-Benoît Claude,[1] Guillaume Baffou,[1] Jérôme Wenger[1,*]

[1] *Aix Marseille Univ, CNRS, Centrale Marseille, Institut Fresnel, 13013 Marseille, France*

[*] *Corresponding author:* jerome.wenger@fresnel.fr

**Abstract**

Plasmonic nanoapertures generate strong field gradients enabling efficient optical trapping of nano-objects. However, because the infrared laser used for trapping is also partly absorbed into the metal leading to Joule heating, plasmonic nano-optical tweezers face the issue of local temperature increase. Here, we develop three independent methods based on molecular fluorescence to quantify the temperature increase induced by a 1064 nm trapping beam focused on single and double nanoholes milled in gold films. We show that the temperature in the nanohole can be increased by 10°C even at the moderate intensities of 2 mW/µm² used for nano-optical trapping. The temperature gain is found to be largely governed by the Ohmic losses into the metal layer, independently of the aperture size, double-nanohole gap or laser polarization. The techniques developed therein can be readily extended to other structures to improve our understanding of nano-optical tweezers and explore heat-controlled chemical reactions in nanoapertures.

**Keywords :** plasmonics, nano-optical trapping, optical tweezers, temperature measurements, nanoaperture, double nanohole, zero-mode waveguide

Plasmonic nanostructures and nanoantennas can confine the light down into sub-wavelength dimensions and locally generate strong field gradients exceeding by far those achievable with conventional diffraction-limited optics.[1] These intense field gradients enable efficient nano-optical trapping of nanoparticles and even single molecules.[2–7] Such unprecedented ability to locate a nanoparticle and/or a quantum emitter with nanometer accuracy opens promising opportunities for many applications in biosciences and quantum information processing.[8–14] Among the different plasmonic nanostructures used for nano-optical trapping, single and double nanoholes milled in gold films have been widely used.[2,3,15–21] The single nanohole structure benefits from a self-induced back action (SIBA) where the presence of the trapped object helps to further improve the restoring force into the trap.[15] Additionally, the double nanohole (and similarly the bowtie aperture) structures benefit from enhanced field gradients at the apex joining the two apertures to further sharpen the trapping potential.[2,17,19,20]

A common feature to all plasmonic devices are the absorption and Ohmic losses into the metal leading to Joule heating of the surrounding environment.[22,23] This has led to the fruitful area of thermoplasmonics, where the heat generation in plasmonic nanostructures is used to control the local temperature distribution and realize heat-induced applications.[24–26] However, in the case of optical trapping, a local temperature increase and a temperature gradient around the trap can lead to unwanted effects affecting the trapping potential and/or the trapped object.[27–29] Therefore, low-loss alternative antennas based on dielectric materials are being explored.[30–33] However, while metal plasmonic nanostructures remain largely used, the question of measuring the local temperature increase in single and double nanoholes remains open. Recently, numerical simulations have been reported for double nanoholes, indicating a temperature increase around 10°C at the typical 10 mW/µm² intensity used for nano-optical trapping.[21] This somehow contradicts the generally believed argument that the temperature increase in the case of nanoholes remains below a few degrees as the metal layer acts as an efficient heat sink to further dissipate the Joule heating.[12,29] Therefore, we believe that a clear experimental investigation is needed to assess the temperature increase in nanohole-based optical tweezers.

Here, we use fluorescence spectroscopy to locally measure the temperature into single and double nanoholes milled in gold films and quantify the temperature increase induced by the 1064 nm trapping beam. We discuss the implementation of three different fluorescence readouts: the fluorescence intensity, diffusion time and fluorescence lifetime. While these readouts are largely independent from each other, we show that they all lead to similar results and can be used in an equivalent way. Our measurements stand in excellent agreement with numerical simulations, and indicate that temperature increases around 10°C can be reached even at the moderate intensities of 2 mW/µm² typically used for nano-optical trapping. We also show that the temperature increase is largely independent of the aperture size, double-nanohole gap and heating laser polarization. In other words, the temperature increase is largely governed by the absorption of the 1064 nm laser into the metal layer, and does not depend on the plasmonic properties of the nanohole. The ability to accurately and remotely control the temperature into nanoholes of sub-femtoliter volumes opens their use for a broad range of biochemical applications from biosensing to crystallization.[28,34–37]

**Theory**

Many different techniques have been developed to probe the temperature at a sub-micron scale, well below the spatial resolution limit of classical thermometers and thermocouples.[26,38–40] Readouts based on fluorescence combine the advantage of high sensitivity, low invasiveness and good spatial resolution (when conducted on a confocal microscope).[38,41] We describe here the underlying physical principles and assumptions behind the three different measurements used experimentally, based on fluorescence lifetime, intensity and diffusion time.

*Fluorescence lifetime.*

For many fluorescent molecules, such as Alexa Fluor 647 used here, a temperature change will affect the non-radiative decay rate, leading to a change in the apparent fluorescence lifetime.[42,43] The fluorescence lifetime $\tau_T$ of Alexa Fluor 647 in a homogeneous water medium at a temperature T can be written as:

$$\frac{1}{\tau_T} = \Gamma_{rad} + \Gamma_{nrad}(T) \qquad (1)$$

where $\Gamma_{rad}$ and $\Gamma_{nrad}$ are the radiative and non-radiative decay rates. Here, we assume that the radiative decay rate is constant with the temperature: $\Gamma_{rad}(T) = \Gamma_{rad}$. The agreement between our different temperature measurements (based on lifetime, intensity and diffusion time) validates this assumption *a posteriori*. By measuring the fluorescence lifetime for a homogeneous solution set at different temperatures (Fig. 2f), and knowing the 33% quantum yield of Alexa Fluor 647 at 21°C, the the non-radiative decay rates $\Gamma_{nrad}(T)$ can be calibrated at different temperatures. A more convenient alternative is to express the temperature increase $\Delta T$ as a function of the change in the non-radiative decay rates:

$$\Delta T = 91.31 - 91.22 \times \left(\frac{\Gamma_{nrad}(21°C)}{\Gamma_{nrad}(T)}\right)^{0.42} \qquad (2)$$

This calibration will be used to compute the temperature increase in nanoholes based on fluorescence lifetime measurements.

The presence of a plasmonic nanostructure affects the fluorescence lifetime by enhancing the radiative decay rate (Purcell effect) and opening new decay pathways to the free electrons in the metal.[23] Within the nanohole, the fluorescence lifetime is changed to $\tau_T^*$, the radiative rate is increased to $\Gamma_{rad}^*$ and an additional nonradiative decay rate $\Gamma_{loss}^*$ is introduced to account for the plasmonic losses due to the presence of the metal structure:

$$\frac{1}{\tau_T^*} = \Gamma_{rad}^* + \Gamma_{loss}^* + \Gamma_{nrad}(T) \qquad (3)$$

In this equation, $\Gamma_{loss}^*$ describes the nonradiative energy transfer to the free electrons in the metal (quenching phenomenon) while $\Gamma_{nrad}(T)$ describes the nonradiative decay internal to the molecule (internal conversion). Our primary interest here is to measure the temperature change induced by the presence of an external infrared laser beam in the nanostructure environment. For a given nanohole size, the presence of a heating laser will increase the temperature and thus the non-radiative decay rate $\Gamma_{nrad}$, leading to an supplementary reduction of the fluorescence lifetime $\tau_T^*$ in the nanohole as compared to the fluorescence lifetime $\tau_{21°C}^*$ still in the nanohole but without the infrared laser heating. We assume here that the radiative rate $\Gamma_{rad}^*$ and the plasmonic loss rate $\Gamma_{loss}^*$ are not noticeably affected by the temperature. Therefore, the temperature increase can be determined solely from the relative change of the non-radiative decay rates inside the nanohole, deduced from the fluorescence lifetimes $\tau_T^*$ and $\tau_{21°C}^*$ measured in the nanohole with and without the heating laser:

$$\frac{\Gamma_{nrad}(T)}{\Gamma_{nrad}(21°C)} = 1 + \frac{1}{\Gamma_{nrad}(21°C)} \left(\frac{1}{\tau_T^*} - \frac{1}{\tau_{21°C}^*}\right) \qquad (4)$$

Combining then equations (2) and (4), we can express the relationship linking the temperature increase $\Delta T$ with the fluorescence lifetimes $\tau_T^*$ and $\tau_{21°C}^*$ in the nanohole (respectively with and without the infrared heating laser):

$$\Delta T = 91.31 - 91.22 \times \left[1 + \frac{1}{0.67}\left(\frac{1}{\tau_T^*} - \frac{1}{\tau_{21°C}^*}\right)\right]^{-0.42} \qquad (5)$$

Here $\tau_T^*$ and $\tau_{21°C}^*$ are expressed in nanoseconds, and we have used the fact that $\Gamma_{nrad}(21°C) = 0.67$ ns$^{-1}$ using the 1.0 ns lifetime and 33% quantum yield of Alexa Fluor 647 at room temperature. The validity of this approach will be demonstrated in the Results and Discussion section by comparing to the results found with the numerical simulations and the two other experimental methods based on the fluorescence intensity and diffusion time.

*Fluorescence intensity.*

As a consequence of the change in the balance between the radiative and non-radiative decay rates, the net quantum yield of the emission process changes with the temperature, leading to a variation of the detected fluorescence intensity.[39,41,44] Other processes may also affect the total detected fluorescence intensity such as the triplet blinking kinetics and photo-isomerization rates which were reported to be temperature-dependent in the case of cyanine5 (an analog to Alexa Fluor 647).[54] All these processes contribute to decrease the net fluorescence intensity detected for Alexa Fluor 647 as the temperature is increased. The temperature dependence of the fluorescence intensity has been calibrated on a spectrofluorometer (Fig. 2a,d). Based on numerical interpolation of this calibration data, we empirically find the following relationship linking the relative temperature increase $\Delta T$ with the fluorescence intensity drop:

$$\Delta T = 25.75 - 25.6 \times \left[\frac{F(\text{T})}{F(21°\text{C})}\right]^{2.2} \qquad (6)$$

While this calibration has been recorded for homogeneous solutions, we apply it also for single and double nanoholes to determine the temperature increase $\Delta T$ from the ratio in the measured fluorescence intensities $F^*(T)/F^*(21°C)$ with and without the infrared heating laser. This is an approximation as the presence of the metallic nanostructure will affect the different decay rates, modifying the apparent quantum yield and also the dark state (triplet) kinetics. As we show below, in the case of the single and double nanoholes used here, the temperatures inferred with this approach are well within the range found with the two other experimental methods and the numerical simulations. The reason why this approach remains valid is that for nanoholes of 200-500 nm diameters as well as double nanoholes with 30-40 nm gap sizes, the modification of the radiative and loss decay rates due to the presence of the plasmonic nanostructure ($\Gamma_{rad}^*$ and $\Gamma_{loss}^*$) remains moderate as compared to the intrinsic nonradiative rate $\Gamma_{nrad}$ of Alexa Fluor 647.[55,56] The situation could be different if a high quantum yield dye was used in conjunction with a highly resonant dimer plasmonic nanoantenna.[57] The main advantage of the approach using Eq. (6) is its simplicity as no advanced fluorescence analysis (lifetime, FCS) is needed. Additionally, we demonstrate in the Supporting Information section S6 that quantitative temperature estimates can be rigorously recovered from the measured fluorescence intensities using the nonradiative rate dependence Eq. (4,5) obtained from lifetime analysis.

*Fluorescence diffusion time.*

Beyond the changes in the non-radiative decay rate leading to a reduction in both the lifetime and the quantum yield, the translational [45] and rotational [46] diffusion of the molecules will also be modified by a temperature change. The molecular diffusion coefficient D depends on the temperature T according to the Stokes-Einstein relationship:

$$D(T) = \frac{kT}{6\pi\eta(T)r} \qquad (7)$$

where D is the diffusion coefficient, k the Boltzmann constant, and r the hydrodynamic radius. The viscosity η of the medium surrounding the molecules also depends on the temperature. Assuming that the viscosity of Alexa Fluor 647 solution with a micromolar concentration is similar to the viscosity of pure water, the temperature dependence of the viscosity can be described as:

$$\eta(T) = A \cdot 10^{B/(T-C)} \qquad (8)$$

where the constants A, B and C are $2.414 \cdot 10^{-5}$ Pa·s, 247.8K and 140K respectively.[58] Fluorescence correlation spectroscopy (FCS) is a well-established technique to probe the molecular translational diffusion and record the mean diffusion time of the fluorescent dyes across the detection volume.[47–50] Since the FCS diffusion time $\tau_d$ is proportional to $1/D(T)$, the change in the FCS diffusion time can then be computed back into a temperature change using the Stokes-Einstein equation (7).[43,45,51–53] Here we neglect the influence of the convection phenomenon as it has been reported that convection flows played a negligible role inside the nanoaperture.[21,59]

Specifically, for any given experimental geometry (confocal, single or double nanohole…), the ratio between the diffusion time without the infrared laser at the room temperature (294.15K, heating laser off) and the one at a certain temperature *T* (in the presence of the heating infrared laser) is given by:

$$\frac{\tau_d(T)}{\tau_d(294.15K)} = \frac{\eta(T) \cdot 294.15}{\eta(294.15) \cdot T} = \frac{10^{B/(T-C)} \cdot 294.15}{10^{B/(294.15-C)} \cdot T} \qquad (9)$$

Hence the absolute temperature *T* can be derived once known the relative change in the measured FCS diffusion times with/without the heating infrared laser. This ratio does not depend much on the model equation used to fit the FCS data (the FCS correlation function width at half-maximum should suffice), the only requirement is that the triplet (or dark state) blinking kinetics can be readily differentiated in the FCS curve from the diffusion process. This is well the case even for the smallest nanoholes investigated here as the triplet blinking time is below 2 μs while the diffusion time is above 30 μs (Supporting Information Fig. S7).

As we discuss below, all three fluorescence-based readouts – intensity, diffusion and lifetime – can be easily implemented with single and double nanoholes to quantitatively record the temperature change induced by the trapping laser beam.

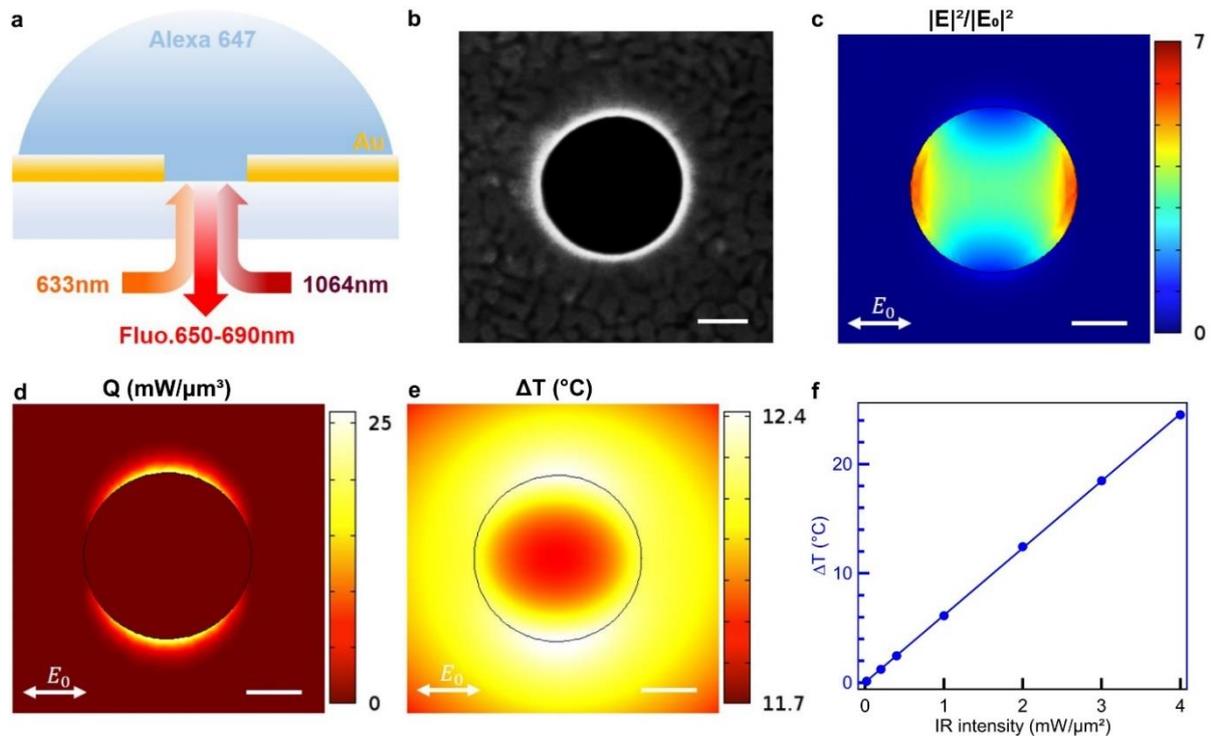

**Figure 1.** Numerical simulations of the temperature increase in a single nanohole in the optical trapping configuration. (a) Scheme of the optical trapping setup with a nanoaperture milled in gold. The 1064 nm trapping beam is overlapped with a 635 nm laser beam to excite the fluorescence of Alexa Fluor 647 molecules filling the nanoaperture volume. (b) SEM image of a single nanohole (SNH) aperture with 300nm diameter. (c)-(e) Simulation predictions of intensity enhancement, heat source density and temperature increase respectively for 2 mW/µm² illumination at 1064 nm from the bottom of the 300 nm diameter SNH. The scale bars in (b-e) are 100 nm and the 1064 nm beam is linearly polarized along the horizontal direction. (f) Numerical simulation of the average temperature increase inside the nanohole as a function of the 1064 nm intensity.

**Results and Discussion**

Our experimental approach combines an optical tweezer with a fluorescence microscope. Single nanohole (SNH) structures milled in gold are set at the focus of a high NA oil immersion objective (Zeiss 40x, NA=1.3) where two laser beams are overlapped: one 1064 nm continuous laser used for optical trapping/heating and one 635 nm pulsed laser used for fluorescence excitation (Fig. 1a). This configuration allows to record the evolution of the fluorescence from Alexa Fluor 647 molecules in water solution as a function of the 1064 nm intensity in conditions similar to the ones commonly used for optical trapping.[1–3,15,16,21] As compared to the Rhodamine B dye widely used for nanothermometry,[41] the choice for a red-emitting cyanine dye such as Alexa Fluor 647 is better suited to preserve the plasmonic properties of the gold film and avoid the metal photoluminescence found for gold layers below 600 nm.[26] Due to the relatively small intensity of the red excitation laser (typically 20 µW/µm²) as compared to the IR intensity (5 mW/µm²), we neglect here the heating effect induced by the red laser and focus on the 1064 nm intensity influence.

A simulation model using COMSOL Multiphysics is first introduced to predict the IR laser heating in a single 300 nm nanohole milled in gold. Maps of the intensity enhancement, heat source density and temperature increase are presented in Fig. 1c-e respectively. The local field intensity enhancement appears maximum at the aperture edges along the illumination polarization where the electrons tend to accumulate. Conversely, the heat source density is higher at the aperture edges in the direction perpendicular to the excitation polarization, where the power dissipated by the light-induced electron current is maximum.[60] This behavior can be intuitively explained by reminding that the heating comes from the velocity motion of the electrons, which is maximum in the middle between the two accumulation points at the aperture edges along the excitation polarization. The Joule heating accumulated around the SNH leads to the temperature increase indicated in Figure 1e. At a 2 mW/µm² IR illumination, an increase of 11.7°C is predicted at the center of the SNH. The slight difference (0.7°C) of temperature between the wall and the center of the SNH can be explained by the diffusive nature of heat that cannot remain as confined as a light field. The temperature decreases with the distance from the nanoaperture, with a typical reduction by 50% at a 2 µm lateral distance (Supporting Information Fig. S1). With such thermal gradients, it has been reported that thermal convection induces a negligible fluid velocity increase below 10 nm/s,[61] and that the fluid velocity should even display a minimum at the nanohole position.[21,59] Therefore, we hereafter neglect the influence of the convection phenomenon. The temperature change with respect to the IR intensity is drawn in Fig. 1f and follows a linear behavior as expected from the linearity of the heat-diffusion equation and the independence of the thermal conductivities over this temperature range.[26] Notably, at 4 mW/µm² IR illumination, the temperature raise in the SNH is predicted to exceed 20°C. This non-negligible value shows that the metal layer is not efficient enough to fully dissipate the heat generated by focused IR illumination.

In order to experimentally verify these numerical predictions, we first calibrate the temperature dependence of Alexa Fluor 647 fluorescence without the presence of the gold film (Fig. 2). These experiments are performed with a heating stage to accurately control the sample temperature and record the evolution of the fluorescence intensity and lifetime as the temperature is gradually increased. The calibration results are summarized in Fig. 2d-f. As expected for common cyanine-based organic fluorophores, Alexa Fluor 647 non-radiative decay rate increases with temperature, inducing a reduction in the fluorescence lifetime, quantum yield and detected brightness (Fig. 2d,f). Simultaneously the viscosity of the medium decreases for higher temperatures, leading to faster molecular diffusion and reduced FCS diffusion times (Fig. 2e). Importantly the curves in Fig. 2d-f allow us to calibrate the evolution of the fluorescence intensity, diffusion time and lifetime (detailed formulas are provided in the Theory section) which will be used to infer the temperature increase in the plasmonic nanoholes under IR illumination. We have also carefully checked that the 1064 nm laser beam had negligible influence on the Alexa 647 fluorescence process with no indication of photodamage or saturation in the range of excitation powers used with nanoholes (Supporting Information Fig. S2). Even at 5 mW/µm² IR power, the 1064 nm CW beam did not generate any detectable two-photon fluorescence from Alexa 647 molecules (Supporting Information Fig. S3).

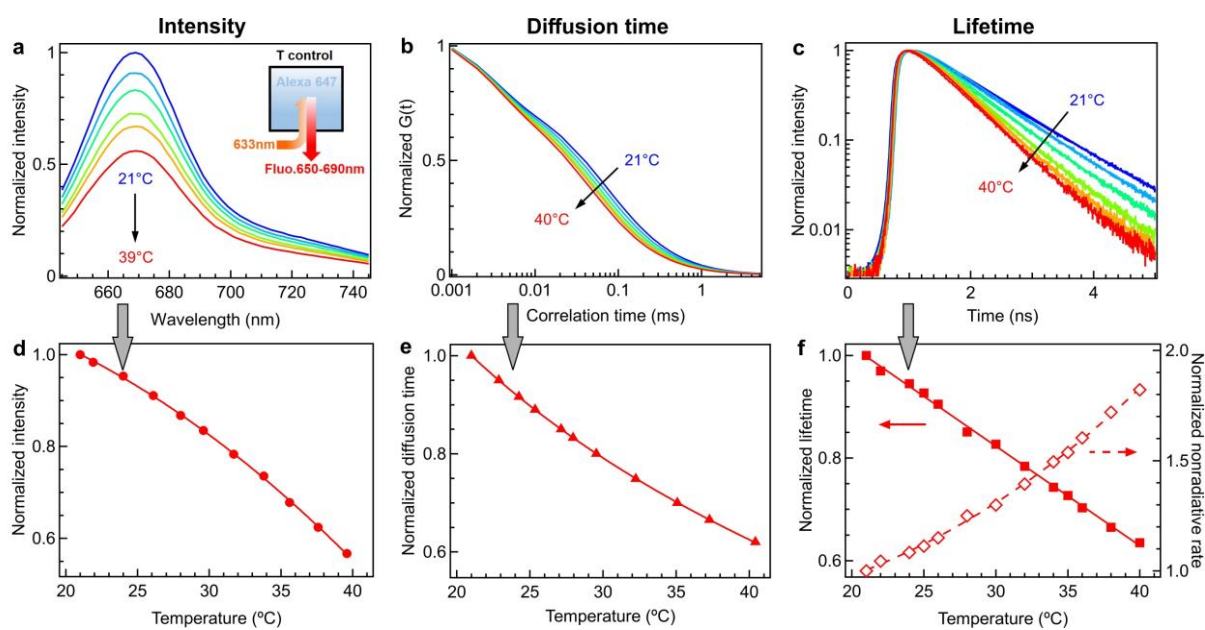

**Figure 2.** Calibration of the Alexa Fluor 647 fluorescence properties as a function of the temperature in water solution. (a) Normalized fluorescence intensity spectra, (b) normalized FCS correlation functions and (c) normalized fluorescence decay traces with increasing temperatures. In (a) the temperatures are respectively 21°C, 26°C, 29°C, 33°C, 35°C and 39°C from top to bottom. In (b) 21°C, 25°C, 30°C, 33°C, 35°C and 40°C from right to left and in (c) 21°C, 26°C, 30°C, 35°C, 38°C and 40°C from top to bottom. From the data set in (a-c), we deduce the calibration curves of the normalized intensity (d), diffusion time (e), lifetime and nonradiative decay rate (f) as a function of the temperature of the solution. The data in (b,e) are calculated based on the Stokes-Einstein equation (7) with the analytical formula of temperature-dependent viscosity of water.[58] The lines in (d,f) are numerical fits.

We next proceed to investigate the temperature increase in single nanoholes under IR illumination. A single nanohole (SNH) milled in 100 nm thick gold with 300 nm diameter is set at the focus of the red and IR lasers and covered with a solution containing Alexa Fluor 647 molecules at 200 nM concentration. It is important to note that the fluorescent molecules are freely diffusing in and out of the nanohole, providing a constant renewal of the fluorescent probes inside the nanohole. Figure 3a shows a typical fluorescence intensity time trace recorded at a constant 20 µW/µm² red excitation intensity while the IR laser with 4mW/µm² intensity is periodically switched on and off. Once the IR laser is on, the fluorescence intensity drops quickly to a stable level before recovering its initial level as the IR laser is turned off. The process if fully reversible and can be repeated for several minutes without any significant change, indicating that fluorophore photobleaching or SNH photodamage do not play any role in our measurements.

For each IR intensity, a 60 s fluorescence time trace is recorded allowing us to compute the FCS correlation function (Fig. 3b) and lifetime decay trace (Fig. 3c). The data recorded with the SNH with increasing IR intensities show the same tendencies as the temperature calibration in Fig. 2. This qualitatively indicates that the IR laser beam induces a local temperature increase in the SNH under optical trapping condition. Quantitatively, we measure the reduction of three fluorescence readouts (intensity, diffusion time and lifetime) with respect to IR intensities (Fig. 3d-f left axis). Using the calibration obtained in Fig. 2, we can then compute back the temperature increase in the SNH as a function of the IR intensity (Fig. 3d-f right axis). While the three readouts remain largely independent from each other, similar temperature increases are assessed with all three methods. Actually the temperature values based on the fluorescence intensity (Fig. 3d) tend to be slightly overestimated by ~1-2°C as compared to the other two methods, as the approach using the calibration Eq. (6) is only an approximation in the case of a plasmonic nanostructure (quantitative measurements are rigorously derived in the Supporting Information section 6). Typically, we observe that the temperature in a 300 nm SNH rises by 20°C under 5 mW/µm² IR illumination, in good agreement with our simulation results (Fig. 1f). While the temperature can also be increased in conventional optical traps due to water absorption,[45,62,63] the temperature rise is typically of 5°C at 100 mW/µm² (Supporting Information Fig. S2). With nanoholes, greater temperature increases can be achieved with significantly less optical power, which can be an additional benefit for temperature-related applications including biochemistry[28] and biosensing.[35–37]

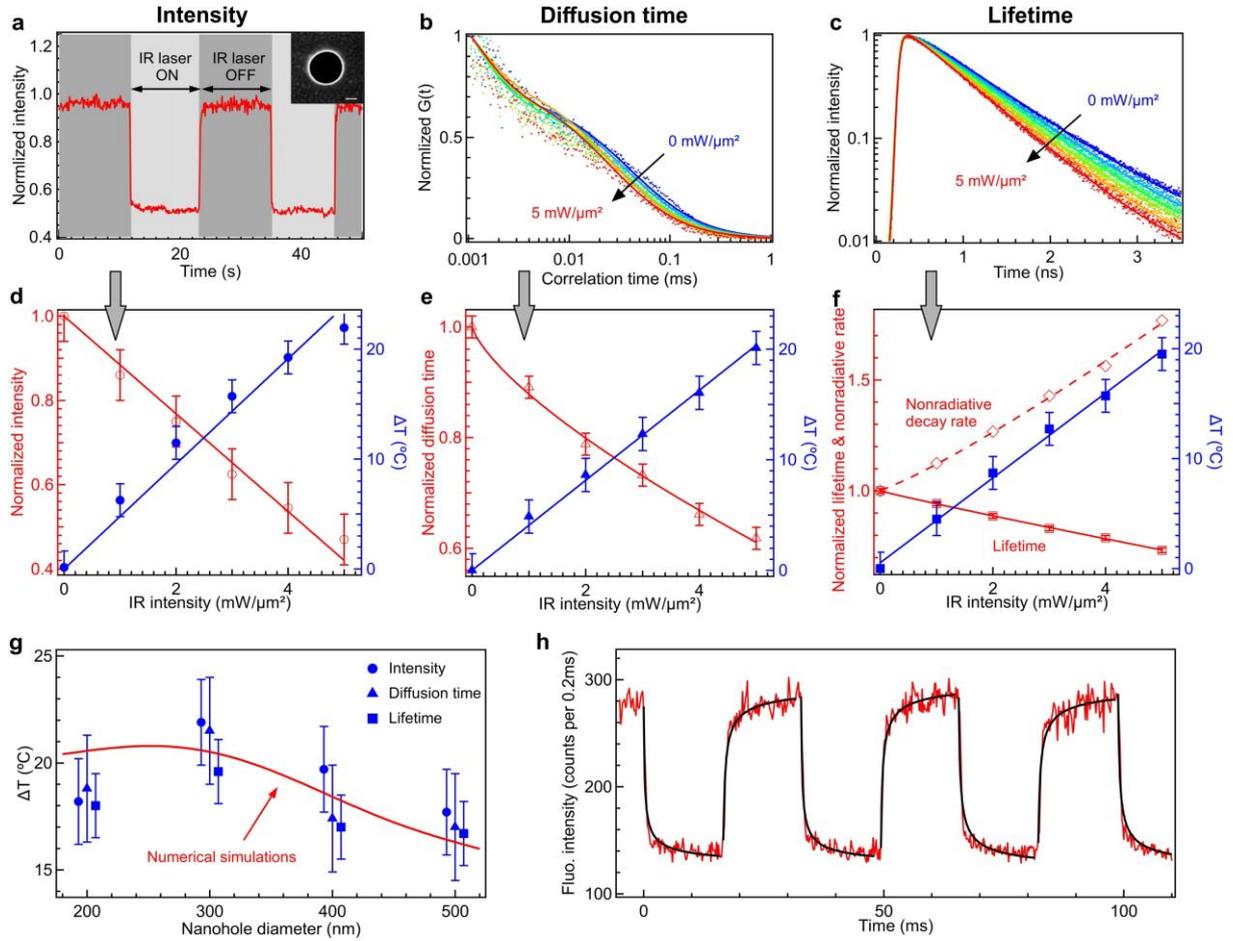

**Figure 3.** Temperature increase in single nanoholes under 1064 nm illumination. (a) Fluorescence intensity time trace with and without 4 mW/μm² illumination at 1064 nm. The fluorescence excitation is constant at 20 μW/μm². The nanohole diameter is 300 nm. (b)-(c) Normalized FCS correlation functions and decay curves recorded for increasing infrared intensities. (d)-(f) Normalized intensity, diffusion time, lifetime and nonradiative decay rate $\Gamma_{nrad}$ of Alexa Fluor 647 solution filling a single 300 nm nanohole as a function of the 1064 nm intensity (red traces, left axis). From the calibration results displayed in Fig. 2, we compute back the temperature increase induced by the infrared laser beam for each measurement: intensity (d), diffusion time (e) and nonradiative rate (f) (blue curves, right axis). Solid lines are guide to the eyes. (g) Comparison of the temperature increase for different nanohole diameters under the same 5 mW/μm² illumination. Markers represent the experimental results from different readouts (circles for intensity, triangle for diffusion time and squares for lifetime, markers are slightly horizontally shifted for clarity). The red line is deduced from the numerical simulations (Supporting Information Fig. S4), and is not a fit to the data. (h) Millisecond dynamics of the fluorescence intensity recorded on a 300 nm nanohole while the infrared intensity is modulated by a chopper blade at 30 Hz. Red lines are experimental data and black lines are numerical fits using error function model $\mathrm{erf}(\sqrt{\tau/t})$.[26] The fluorescence binning time is 200 μs.

The experimental errors for the measurements of intensities, diffusion times and lifetimes mainly stem from the small mechanical drifts of the laser focus and the uncertainties induced by the noise during the curve-fitting process. As the fluorescence lifetime does not depend much on the detected intensity and laser focus, this method appears relatively more accurate than the two other readouts. However, the calibration formula (Eq. (5)) requires the knowledge of the reference quantum yield and lifetime of Alexa 647 at 21°C. We have found that changing the reference quantum yield of Alexa 647 at 21°C form 33% to 30% induced a modification in the measured temperature by 0.5°C, and should be accounted for in the possible experimental uncertainties. On the contrary, a 5% uncertainty on the reference lifetime of Alexa 647 at 21°C does not lead to any significant effect on the final measured temperature (uncertainty below 0.1°C). While comparing the different techniques, it should also be mentioned that the fluorescence lifetime implementation is comparatively more expensive (need for pulsed laser and time-correlated single photon counting electronics). It is interesting to note that a readout based only on the total detected fluorescence intensity (Fig. 3d) is very simple to implement and readily provides an estimate of the temperature increase.

Our numerical simulations do not indicate any major change of the temperature increase with the nanohole diameter for diameters below 350 nm (Supporting Information Fig. S4). For this range of diameters, the temperature increase is essentially the same than for a continuous gold film. For diameters above 350 nm, the temperature elevation starts to go down with the SNH diameter as the infrared light transmission through the SNH increases. We verify these predictions by measuring SNHs of different diameters (similar to the ones used for SIBA-trapping[15]) under similar illumination conditions (Fig. 3g). For SNH diameters ranging from 200 to 500 nm, our experimental results indicate a temperature change below 15%, in good agreement with the numerical simulations. Altogether, these results indicate that the heating effect in the SNH is primarily a consequence of the absorption and Joule losses into the metal layer, which do not significantly depend on the transmission or resonance properties of individual nanoholes for diameters below 350 nm. As compared to the 1 µm² illumination area, the SNH surface accounts for less than 10% for diameters below 350 nm (Supporting Information Fig. S5), which explains in part the relative independence of the heating with the SNH diameter. Looking into more details, the numerical simulations and our experimental observations indicate that 300 nm is the largest SNH diameter to maximize the heating, suggesting that for this diameter the field enhancement for the infrared laser compensates for the smaller metal surface illuminated. Interestingly, 300 nm corresponds to the cut-off diameter for the transmission at 1064 nm,[64] and was also reported to be the optimum diameter for SIBA trapping.[15]

We next focus on the temporal dynamics of the fluorescence signal (and the temperature) on a millisecond timescale. To this end, the 1064 nm laser beam is modulated at 30 Hz with a rotating chopper blade. Figure 3h shows a typical fluorescence intensity time trace similar to the one in Fig. 3a but with faster dynamics and better temporal resolution aiming to estimate the heating and cooling times. The transient evolution is well described by a model based on the error function $\text{erf}(\sqrt{\tau/t})$ where τ is a constant describing the heating/cooling timescale and t denotes the time variable.[26] Interestingly, the transient evolution of the fluorescence intensity cannot be well fitted by a simple exponential profile, one has to use the error function, as theoretically derived in the case of transient heating and cooling.[26] This further confirms that the fluorescence drop originates from a local temperature change. The numerical interpolation of the experimental data estimates the heating time $\tau_{heat}$ = 90 ± 40 μs and the cooling time $\tau_{cool}$ = 130 ± 30 μs. Both heating and cooling times appear thus on comparable timescales in the hundred microsecond range. We note that despite the data interpolation, these measurements remain limited by the minimum 200 μs binning time required to accumulate enough photon counts without saturating the avalanche photodiodes. Therefore, the transient dynamics of the temperature may even occur on a faster timescale.

Having quantified the temperature increase in a single nanohole, we now investigate the double nanohole (DNH). The DNH apertures are milled in the same gold film as the SNHs, and comprise two circular holes of 200 nm diameter connected by a 80 nm long groove. Two gap sizes (the width of the groove) are used in the experiments with 30 nm and 40 nm, as measured by the SEM images (Fig. 4a). These structures correspond to the ones typically used in nano-optical tweezers,[2,3,18,21] and are also representative of bowtie nanoapertures.[16,17,20] Numerical simulations for the field intensity enhancement and temperature increase for a 40 nm gap DNH are displayed on Fig. 4a,b. When the IR laser polarization is set parallel to the metal apex between the holes, the intensity enhancement reaches a maximum while it nearly vanishes when the polarization is turned by 90° to be aligned with the main axis joining the two holes (Fig. 4b). This behavior confirms that the DNH structure with 40 nm gap follows the expected trend for a plasmonic antenna.[2,56] Interestingly, the temperature increase in the DNH follows a completely different behavior with the laser polarization orientation (Fig. 4b). Although the intensity enhancement displays a huge difference between 0° and 90°, the temperature increase remains nearly constant around 8.5°C under 2 mW/μm² IR laser illumination, suggesting that the plasmonic response of the gap mode in the DNH plays almost no role in the heat generation.

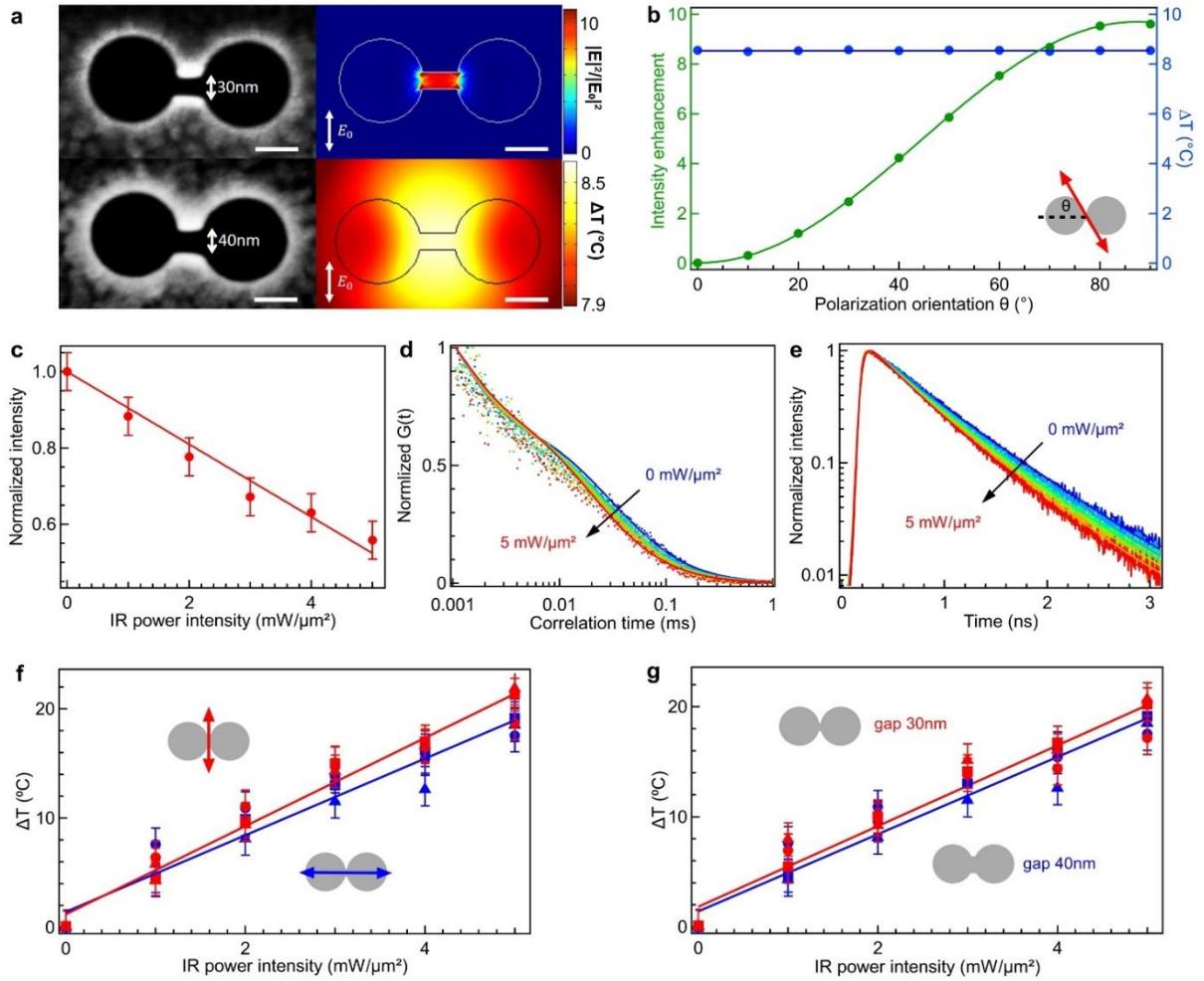

**Figure 4.** Temperature increase recorded on double nanoholes under 1064 nm illumination. (a) SEM images of DNHs with 30 and 40 nm gaps. The right-top and the right-bottom images are the simulation predictions for the intensity enhancement and the temperature increase at 2 mW/µm² IR illumination from the bottom of the DNH with 40nm gap. The polarization of the IR laser is fixed vertically as indicated by the arrows. All the scale bars are 100nm. (b) Local excitation intensity enhancement at 1064 nm (green, left axis) and temperature increase (blue, right axis) as a function of the polarization orientation θ. (c)-(e) Normalized fluorescence intensities, correlation functions and fluorescence decay curves measured for the DNH of 40 nm gap with increasing IR intensities. (f) Temperature increase estimated from the fluorescence intensity (circles), diffusion time (triangles) and lifetime (squares) for increasing 1064 nm intensities with 2 different polarizations (red parallel to the apex, blue perpendicular) and 40 nm gap. (g) Same as (f) for two different gap sizes and polarization parallel to the apex. For all these experiments, the 635 nm laser polarization used to excite the fluorescence is set parallel to the apex between the holes (θ = 90°).

To verify the simulations predictions, we measure the temperature increase in the DNH using all three fluorescence readouts. Figure 4c-e show the raw experimental data. Similarly, we convert the three readouts to the temperature increases based on the calibration from Fig. 2d-f. Experimental data points found for parallel and perpendicular polarization overlap with each other with the tolerance of the experimental uncertainties (Fig. 4f). This indicates that the temperature increase does not depend on the incident polarization, in excellent agreement with the numerical simulations (Fig. 4b) and other independent measurements on bowtie nanoapertures.[37] We also probe two different gap sizes and find again similar temperature increases (Fig. 4g). This experimental finding confirms the trend observed for the SNH where the temperature was nearly independent of the nanohole diameter (Fig. 3g). We also find that SNH and DNH structures feature similar temperature gains for the same IR illumination intensity (typically ~8°C at 2 mW/µm²). The temperature increase for a bowtie aperture milled in a 100 nm thick gold film was recently measured at 3.6°C for 3.75 mW/µm² intensity at the objective focus spot.[37] This value is lower than the one that we measure experimentally, we ascribe this difference to the presence of the titanium adhesion layer in our case, while no adhesion layer was used in ref.[37] Interestingly, this observation suggests new routes to tune the local temperature increase, which we are currently exploring. It should still be noted that the temperature increase in the presence of a continuous metal film [29] remains always significantly lower than for isolated structures such as nanoparticles[25] or nanoantennas.[35] Altogether, our results emphasize that the temperature increase is mainly ascribed to the IR Ohmic losses into the metal layer and does not depend on the plasmonic properties of the aperture structures.

**Conclusions**

We have developed three independent methods based on molecular fluorescence to quantify the local temperature increase occurring in single and double nanoholes illuminated by a focused infrared laser beam. Our setup reproduces the configuration and laser power typically used for nano-optical trapping experiments. We show that despite the presence of the gold metal layer acting as a heat sink, the local temperature in the nanohole can be increased by several degrees Celsius even at the moderate mW/µm² infrared laser intensities. In excellent agreement with numerical simulations, we show that the local temperature increase is mainly due to the Ohmic losses into the metal layer and does not significantly depend on the geometry of the nanohole structure. We also discuss the use of three independent fluorescence readouts based on the fluorescence intensity, diffusion time and lifetime. While they all lead to similar results, the fluorescence intensity of Alexa Fluor 647 molecules is certainly the easiest and cheapest method to implement to obtain a first estimate of the local temperature without requiring advanced temporal analysis of the fluorescence emission statistics. The fastest time resolution to investigate temperature kinetics using our photon-counting technique is around 50 µs, and is set by the minimum time needed to record 100 photons at a 2 Mcounts/s detection rate just below the photodiode saturation level. Altogether, our results are relevant to better understand nano-optical trapping experiments.[3,15] They also open the way to use nanoholes as heat-controlled nano-reaction chambers for specific chemical applications including controlled nucleation and polymerization, crystal growth and phase separation.[28] For biosensing applications, the thermal gradients can improve the translocation through nanopores[35–37] or further promote trapping.[14,65] For single molecule analysis or FCS, the nanoholes add the possibility to easily and locally bring the temperature to the physiological 37°C temperature, which appears especially interesting for the study of lipid or protein species bound to the plasma cell membrane.[66]

**Methods**

*Sample fabrication*

The samples are prepared on 0.15 mm thick borosilicate glass coverslips (refractive index 1.52). First, a 5 nm titanium layer is deposited by electron-beam evaporation to serve as adhesion layer for the gold film. Then, a 100 nm thick gold layer is deposited by thermal evaporation (Oerlikon Leybold Univex 350). Finally, the single and double nanohole structures are milled by gallium-based focused ion beam (FEI dual beam DB235 Strata) using 30 kV energy and 10 pA beam current. We have not observed any photodamage to our nanostructures for the 0-5 mW/µm² CW intensities at 1064 nm used here (the damage threshold for gold films is expected to be around 50 mW/µm² at 1064 nm).

*Numerical simulations*

"Electromagnetic Waves, Beam Envelopes" and "Heat Transfer in Solids" modules of COMSOL Multiphysics are used and coupled via the "Electromagnetic Heat Source" module in 3D geometry. The thermal parameters for glass, titanium, gold and water are taken from the COMSOL material library and assumed to be temperature independent. The complex refractive index for Au (100nm thick) and Ti (10nm thick) are taken from Johnson and Christy[67] for bulk materials (i.e. $n_{Au}$=0.25846 + i6.9654, $n_{Ti}$=3.4654 + i4.0085), the refractive index of water and glass are taken as 1.33 and 1.52 respectively. The incident laser beam at 1064 nm follows a Gaussian amplitude profile of 1 µm waist with linear polarization. In the case of a focused beam on a continuous thin film, the polarization is not expected to have a significant influence on the absorption by the thin film (Supporting Information Fig. S6).

*Experimental setup*

The experimental setup is based on an inverted confocal microscope featuring two overlapping excitation beams: one continuous wave (CW) 1064 nm laser (Ventus 1064-2W) and one 635 nm pulsed laser diode (Picoquant LDH-P-635) at 80 MHz repetition rate. The infrared beam is used for optical trapping and heating while the red laser is used to excite the fluorescence of Alexa Fluor 647 molecules. Typically, the 1064 nm laser power is in the mW range while the 635 nm laser power is in the µW range. Both lasers are focused by a high NA microscope objective (Zeiss Plan-Neofluar 40x, NA 1.3, oil immersion). By recording images of SNH with 200 nm diameter, we estimate that the laser spot sizes at the objective focus are 1 µm and 0.6 µm in diameter for the 1064 and 635 nm lasers respectively. The illumination power is expressed in intensity units (mW/µm²) at the objective focus. Hence, our values take into account the 50% transmission of the objective at 1064 nm (information provided by the supplier).

The same microscope objective is used to collect the fluorescence, which is filtered from backscattered laser light by a set of dichroic mirrors, long pass filters, 30 µm confocal pinhole and bandpass filters. Two avalanche photodiodes (Picoquant MPD-5CTC) separated by a 50/50 beam-splitter record the fluorescence photons in the 650-690 nm spectral range and avoid the afterpulsing issue in FCS. The fast timing output of the photodiodes is then sent to a time correlated single photon counting (TCSPC) electronic module (Picoquant Picoharp 300 with PHR 800 router) with time-tagged time-resolved (TTTR) option. The overall temporal resolution of our setup (full-width at half maximum of the instrument response function) is 100 ps. All fluorescence time traces are analyzed with the Symphotime 64 software (Picoquant) enabling to compute the intensity time trace, FCS correlation function and TCSPC decay histogram. For the calibration experiments in Fig. 2b,c, a heating stage (CherryTemp) was set in contact with the 30 µL Alexa solution sample to control its temperature. For the temperature-dependent fluorescence emission spectra experiment, we used an automated cuvette spectrophotometer (Tecan Spark 10M).

*FCS analysis*

We fit the FCS correlation data using a three dimensional Brownian diffusion model with an additional blinking term:[48]

$$G(\tau) = \frac{1}{N}\left[1 + \frac{T_{ds}}{1-T_{ds}} \exp\left(-\frac{\tau}{\tau_{ds}}\right)\right]\left(1+\frac{\tau}{\tau_d}\right)^{-1}\left(1+\frac{1}{\kappa^2}\frac{\tau}{\tau_d}\right)^{-0.5} \qquad (10)$$

where N is the total number of molecules, $T_{ds}$ the fraction of dyes in the dark state, $\tau_{ds}$ the dark state blinking time, $\tau_d$ the mean diffusion time and $\kappa$ the aspect ratio of the axial to transversal dimensions of the nanohole volume. While the assumption of free 3D diffusion is obviously not truly fulfilled in the nanoholes, the above model equation was found to empirically describe well the FCS data with single and double nanoholes, provided that the aspect ratio constant is set to $\kappa$ = 1 as found previously.[55,56] The Supporting Information Fig. S7 shows representative FCS correlation functions and their numerical fits for 300 nm nanoholes at different infrared intensities. The diffusion time $\tau_d$ and the dark state blinking time $\tau_{ds}$ always differ by a factor greater than 30×, enabling the straightforward distinction between the blinking and translational diffusion contributions in the FCS correlation.

## ASSOCIATED CONTENT

**Supporting Information**

Thermal gradient for a 300 nm nanohole, infrared laser influence on Alexa 647 photophysics, nanohole diameter influence on the temperature increase, incidence angle and polarization influence on the absorption by a continuous gold film, triplet and isomerization kinetics in nanoholes, quantitative temperature estimates using the measured fluorescence intensities and the nonradiative rate calibration.

The Supporting Information is available free of charge on the ACS Publications website at DOI: xxxxxxxxx

**Acknowledgments**

The authors thank Franck Thibaudau for stimulating discussions, Mathieu Mivelle for help with the deposition of the gold film and Johann Berthelot for contributing to the early development of the experimental setup.

**Funding Sources**

This project has received funding from the European Research Council (ERC) under the European Union's Horizon 2020 research and innovation programme (grant agreement No 723241 TryptoBoost) and from the Agence Nationale de la Recherche (ANR) under grant agreement ANR-17-CE09-0026-01 and ANR-18-CE42-0013.

**Conflict of Interest**

The authors declare no competing financial interest.

# Supporting Information for

# Temperature Measurement in Plasmonic Nanoapertures used for Optical Trapping

Quanbo Jiang,[1] Benoît Rogez,[1] Jean-Benoît Claude,[1] Guillaume Baffou,[1] Jérôme Wenger[1]

[1] Aix Marseille Univ, CNRS, Centrale Marseille, Institut Fresnel, 13013 Marseille, France

This document contains the following supporting information:

    S1. Thermal gradient for a 300 nm nanohole

    S2. Infrared laser influence on Alexa 647 photophysics

    S3. Nanohole diameter influence on the temperature increase

    S4. Incidence angle and polarization influence on the absorption by a continuous gold film

    S5. Triplet and isomerization kinetics in nanoholes

    S6. Quantitative temperature estimates using the measured fluorescence intensities and the nonradiative rate calibration

**S1. Thermal gradient for a 300 nm nanohole**

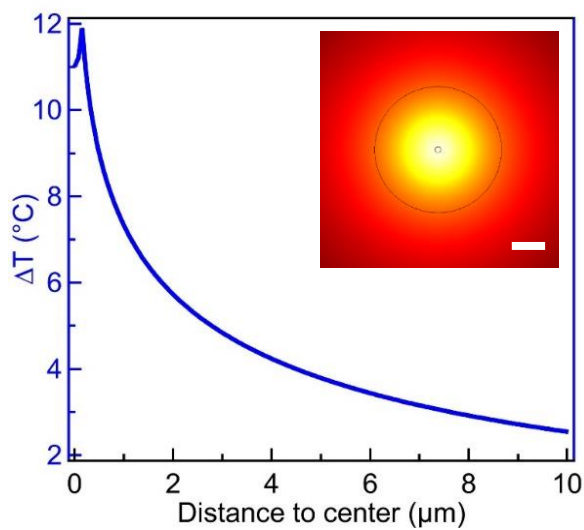

**Figure S1.** Numerical simulation of the evolution of the temperature increase with the distance from the center of a 300nm single nanohole under 2 mW/µm² of infrared illumination. The temperature increase is reduced by 50% at 2 µm from the center of the nanohole. The inset image shows the temperature increase map on a larger spatial scale, the scale bar is 1 µm.

**S2. Infrared laser influence on Alexa 647 photophysics**

Here we investigate the potential influence of the 1064 nm illumination on the Alexa Fluor 647 sample in water solution (without the nanoapertures or metal layer). Figure S2 summarizes our main results. It was shown earlier that CW illumination at 1064 nm of water lead to a local temperature increase,[1] and we retrieve here similar values. However, one has to use intense IR power in order to observe a significant change in the fluorescence photophysics of Alexa Fluor 647. Even at the IR intensity of 80 mW/μm$^2$, the fluorescence intensity, diffusion time and lifetime are only reduced by a factor 0.88 (Fig. S2d-f). In the regime below 5mW/μm$^2$ used in the main article, the 1064 nm illumination induces a reduction below 2% in the absence of the metal layer. This means that despite relatively intense IR intensity, the absorption of Alexa 647 is nearly insensitive to the 1064 nm radiation.

Additionally, we check that the 1064 nm laser does not induce any fluorescence emission from Alexa 647 with two-photon excitation (Fig S3a), and that even under 5mW/μm² infrared illumination the fluorescence intensity of Alexa 647 scales linearly with the 635 nm excitation power (Fig. S3b), showing no sign of saturation or photobleaching.

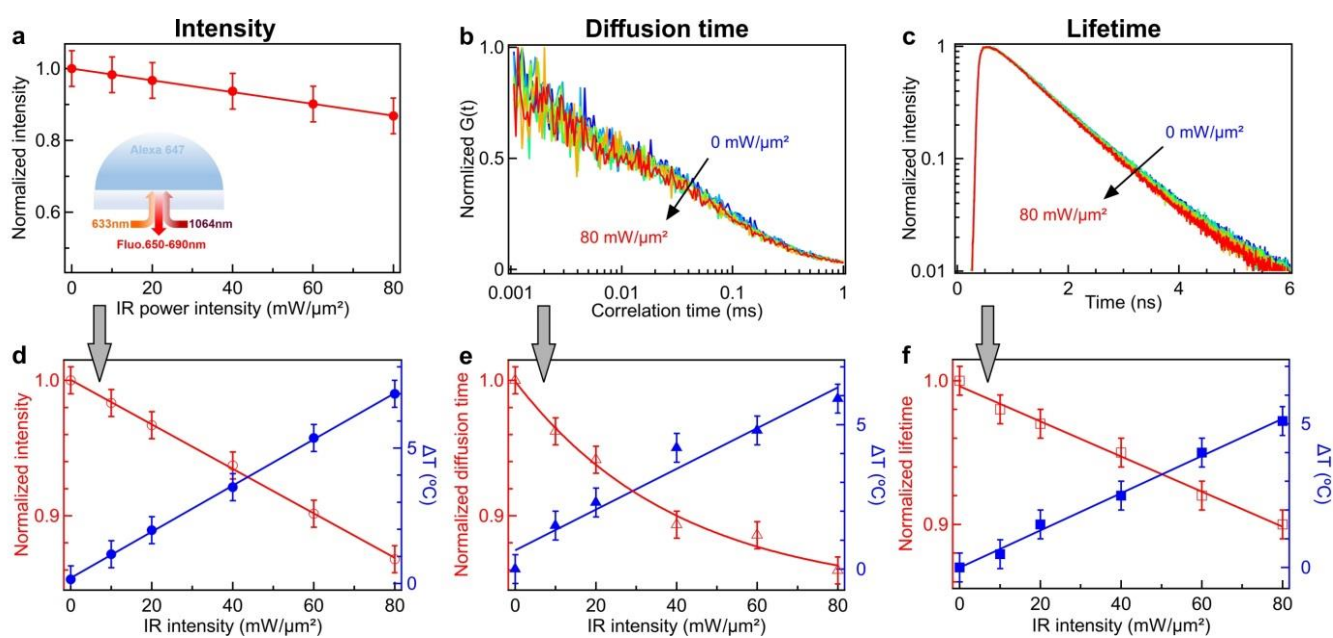

**Figure S2.** (a) Normalized fluorescence intensities of Alexa Fluor 647 solution with 6 different IR intensities in the confocal volume. A droplet of 40 μL Alexa Fluor 647 solution is deposited directly on the substrate without metal layers. (b)-(c) Normalized FCS correlation function and fluorescence decay curves of Alexa Fluor 647 solution with 6 different IR intensities in the confocal volume. (d)-(f) Normalized intensity, diffusion time and lifetime of Alexa Fluor 647 solution calculated from (a)-(c) at the 6 different IR power intensity shown as the red empty round, triangle and square stickers respectively. The red lines are the fits for each result. According to the temperature dependence from the calibration in Fig. 2d-f, the temperature increase based on the fluorescence intensity, diffusion time and lifetime are determined as the blue solid round, triangle and square markers respectively. The blue lines are the fits for each result.

---

[1] Ito, S.; Sugiyama, T.; Toitani, N.; Katayama, G.; Miyasaka, H. Application of Fluorescence Correlation Spectroscopy to the Measurement of Local Temperature in Solutions under Optical Trapping Condition. J. Phys. Chem. B 2007, 111 (9), 2365–2371.

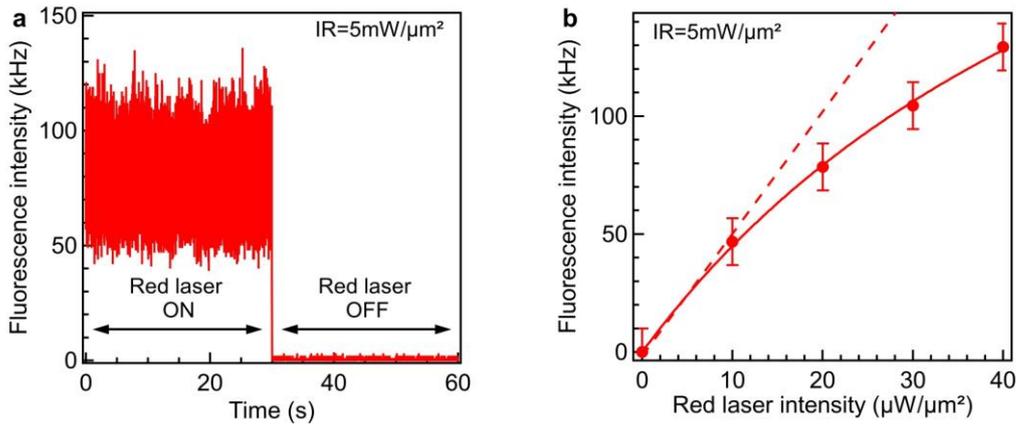

**Figure S3.** (a) Time trace of fluorescence intensity of Alexa Fluor 647 solution under 5 mW/µm² IR illumination in the single nanohole (SNH) with 300nm diameter. From 0-30s, the 20 µW/µm² red laser is kept on. After 30s, the red laser is switched off to show that no fluorescence is detected anymore, even with intense 1064 nm illumination. (b) Fluorescence intensity of Alexa Fluor 647 solution as a function of the 635 nm laser intensity. In this experiment performed with a single nanohole of 300 nm diameter, a 5 mW/µm² IR illumination was kept constant.

**S3. Nanohole diameter influence on the temperature increase**

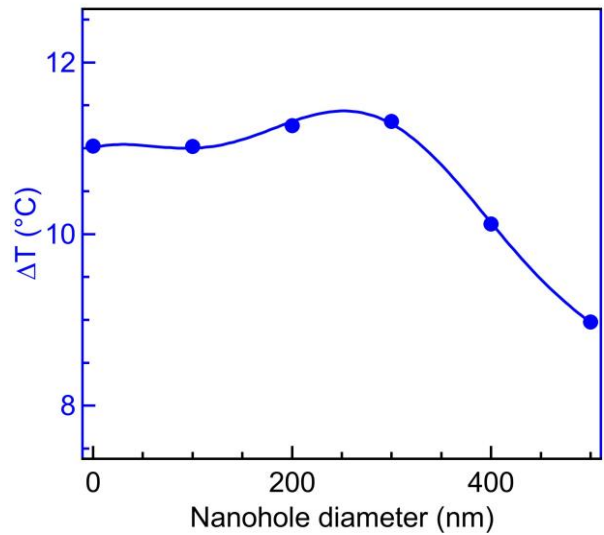

**Figure S4.** Numerical simulations of the temperature increase inside a single nanohole as a function of the nanohole diameter. The infrared intensity is constant at 2 mW/µm² with a beam waist of 1 µm². The temperature remains almost constant when the nanohole diameters are below 350nm and similar to the case of a continuous gold film. For diameters above 350 nm, the temperature elevation goes down with the SNH diameter as the infrared light starts to be transmitted through the SNH and less metal is illuminated.

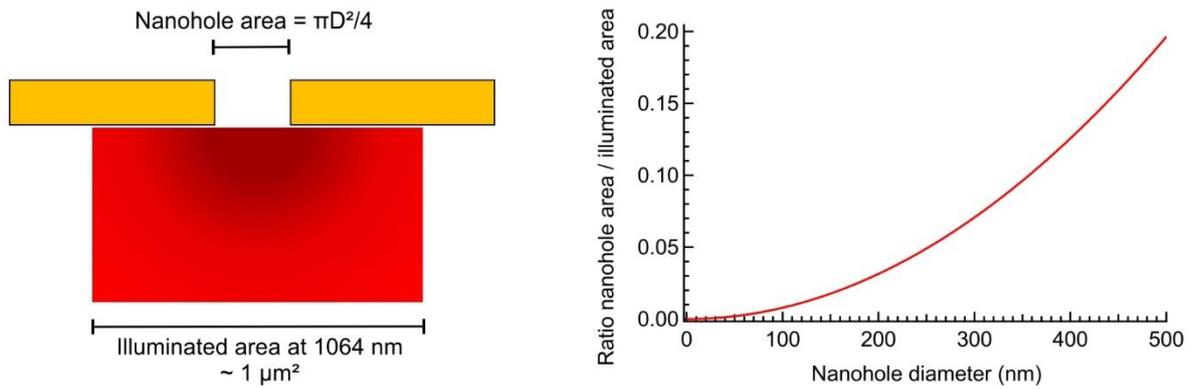

**Figure S5.** Calculation of the ratio of the nanohole geometric area divided by the total illuminated area used experimentally. Nanoholes with diameters below 350 nm account for only a few percents of the illuminated gold surface, which in part explains the relative independence of the temperature increase with the SNH diameter (Fig. S4).

**S4. Incidence angle and polarization influence on the absorption by a continuous gold film**

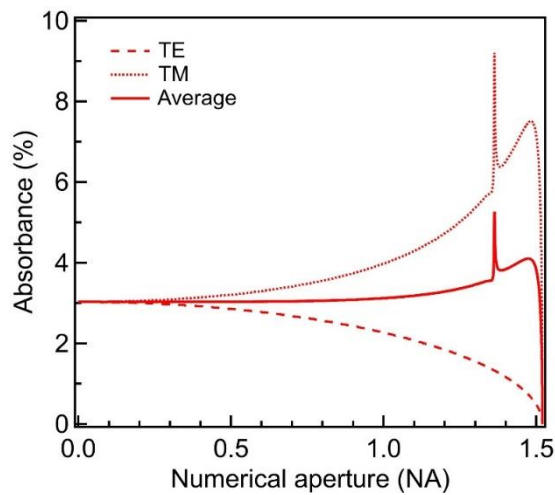

**Figure S6.** Absorbance of the TE and TM polarization in the gold film with respect to the incident angles. The solid curve is the average value of TE and TM modes. The surface plasmon at the gold-water interface is only excited at 1.36 NA which is larger than our 1.3 objective NA.

## S5. Triplet and isomerization kinetics in nanoholes

In addition to the FCS data shown in Fig. 3b, we detail here FCS functions and their numerical fits for time ranges from 100 ns to 10 ms. The use of two avalanche photodiodes as well as FLCS background correction ensures that the phenomenon of afterpulsing does not affect this FCS data and that accurate information about triplet blinking and cis-trans isomerization can be retrieved. The FCS data in Fig. S7 shows that the fast blinking and isomerization kinetics can be readily distinguished from the diffusion contribution in the FCS traces. It also shows that the dark state (triplet) population fraction $T_{ds}$ in the case of the SNH is increased by 30% when the temperature is increased by ~20°C (Fig. S7b,d). Simultaneously, we also monitor a decrease of the triplet blinking time by 60%, which scales with the diffusion time and the viscosity of the water medium. Similar observations were reported while performing temperature-dependent FCS measurements on an Alexa 647 analog,[2] which further confirms the validity of our measurements.

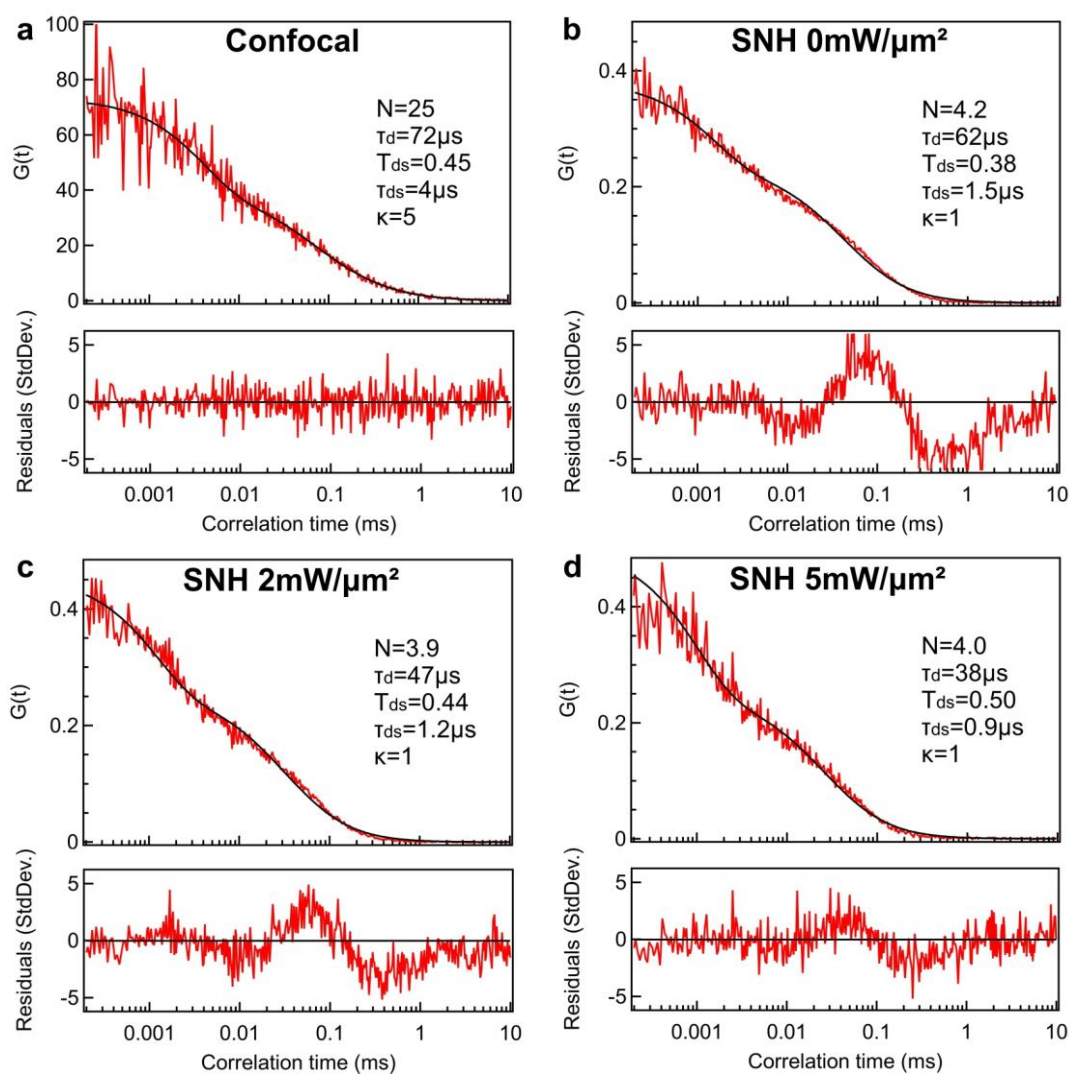

**Figure S7.** FCS correlation (red curves) and numerical fits (black) following the model in Eq. (10) for the confocal reference and a 300 nm gold nanohole at different infrared intensities. The fit parameters are described for each panel. The concentration for the confocal experiment was diluted 10 times.

---

[2] Widengren, J.; Schwille, P. Characterization of Photoinduced Isomerization and Back-Isomerization of the Cyanine Dye Cy5 by Fluorescence Correlation Spectroscopy. J. Phys. Chem. A 2000, 104, 6416–6428.

## S6. Quantitative temperature estimates using the measured fluorescence intensities and the nonradiative rate calibration

The variation in the nonradiative rate induced by the temperature (Eq. 2,5) tends to lower the quantum yield and the fluorescence intensity. Thus quantitative temperature measurements should be achieved using the measured fluorescence intensities based only on the nonradiative rate calibration (Eq. 2,5), i.e. not relying on the empirical formula relating fluorescence intensity and temperature in a homogeneous medium (Eq. 6). This is especially relevant in the case of plasmonic nanostructures which are known to modify the radiative and nonradiative decay rates, thus affecting the apparent quantum yield of fluorescence.

The detected fluorescence intensity $F$ depends on the quantum yield $\phi = \Gamma_{rad}/\Gamma_{tot} = \Gamma_{rad}\, \tau$ (where $\tau = 1/\Gamma_{tot}$ is the fluorescence lifetime) and also on additional terms such as the collection efficiency $\eta$, the number of molecules in the bright state (given by $(1 - T_{ds})N$ as measured from FCS, see Eq (10)) and the excitation intensity $I_{ex}$:

$$F = \eta\, \phi\, (1 - T_{ds})N\, \frac{\sigma\, I_{ex}}{1 + I_{ex}/I_{sat}} \tag{S1}$$

where $\sigma$ is the excitation cross-section (which we assume to be temperature-independent) and $I_{sat}$ is the saturation intensity (which depends on the temperature as it is related to the triplet photokinetics). To relate the variations of the fluorescence intensity $F$ with the change of the quantum yield $\phi$ with the temperature, it is important to take the additional terms also into account (which means that at least an additional FCS analysis is needed such as in Fig. S7):

$$\frac{F(T)}{F(21°C)} = \frac{\phi(T)}{\phi(21°C)} \frac{1 - T_{ds}(T)}{1 - T_{ds}(21°C)} \frac{N(T)}{N(21°C)} \frac{1 + I_{ex}/I_{sat}(21°C)}{1 + I_{ex}/I_{sat}(T)} \tag{S2}$$

Assuming that the radiative decay rate $\Gamma_{rad}$ is independent of the temperature (within the temperature ranges probed here), the ratio of quantum yields $\frac{\phi(T)}{\phi(21°C)}$ is then equivalent to the ratio of fluorescence lifetimes $\frac{\tau(T)}{\tau(21°C)}$. After a bit of algebra, this ratio can be expressed as:

$$\frac{\phi(T)}{\phi(21°C)} = \frac{1}{1 + [\Gamma_{nrad}(T) - \Gamma_{nrad}(21°C)] \cdot \tau(21°C)} \tag{S3}$$

and its temperature dependence is then fully given by Eq. (2): $\Delta T = 91.31 - 91.22 \times \left(\Gamma_{nrad}(21°C)/\Gamma_{nrad}(T)\right)^{0.42}$.

In the presence of a plasmonic nanostructure, it can be shown that without loss of generality the following equation holds:

$$\frac{\phi^*(T)}{\phi^*(21°C)} = \frac{1}{1 + [\Gamma_{nrad}(T) - \Gamma_{nrad}(21°C)] \cdot \tau^*(21°C)} \tag{S4}$$

Here we use the superscript * to indicate that the corresponding variable is taken in the presence of the plasmonic nanostructure. The link connecting the ratio of quantum yields in the nanohole to its counterpart in a homogeneous medium can be written explicitly:

$$\frac{\phi^*(T)}{\phi^*(21°C)} = \frac{1}{1 + \left[\frac{\Gamma_{nrad}(T)}{\Gamma_{nrad}(21°C)} - 1\right] \cdot [1 - \phi(21°C)] \cdot \frac{\tau^*(21°C)}{\tau(21°C)}} \quad (S5)$$

In this equation (S5), the temperature influence acting on the quantum yield in nanohole is entirely given by the ratio $\frac{\Gamma_{nrad}(T)}{\Gamma_{nrad}(21°C)}$ (calibrated from lifetime) and the additional knowledge of the initial 33% quantum yield of Alexa 647 at room temperature and the lifetime reduction $\frac{\tau^*(21°C)}{\tau(21°C)}$ due to the presence of the nanostructure at room temperature. This lifetime ratio amount to 0.72 for Alexa 647 in a 300 nm nanohole milled in a gold film, in agreement with earlier independent measurements.[3]

Combining the equations (S2) and (S5) with the additional quantification of the FCS parameters (Fig. S7), we have everything to compute back the temperature increase based on the measured fluorescence intensities. This approach bridges the gap between the intensity and lifetime methods, and does not rely on the intensity calibration Eq. (6). Only the nonradiative rate calibration with temperature (Eq. (2)) is used here.

Figure S8 summarizes our results and compares between the temperature measurements based on the different approaches. The results based on rigorous analysis of the fluorescence intensity (red disks, Eq. S2 and S5) agree remarkably well with the lifetime measurements (blue squares, same as Fig. 3f) for all infrared intensities. Interestingly, the results derived using the simple intensity calibration (red circles, Eq. 6) tend to overestimate the temperature by ~2°C, yet this appears as a remarkable result given the extreme simplicity of this approach.

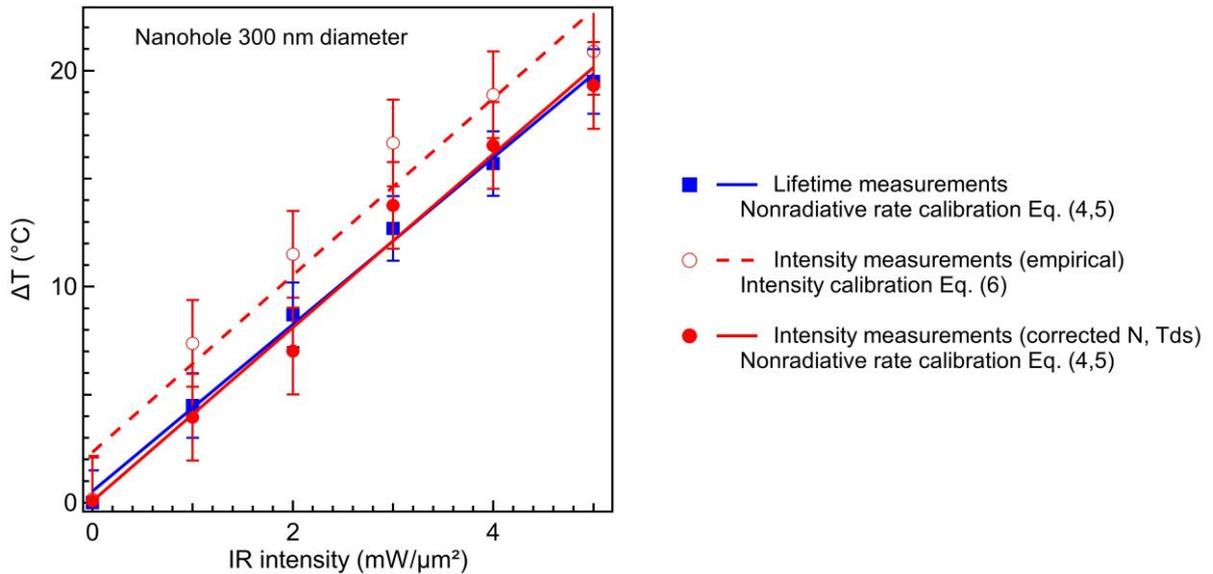

**Figure S8.** Temperature increase $\Delta T$ for different infrared intensities measured from the lifetime data and the nonradiative rate calibration (Eq. 5), using the empirical approach calibrated from the intensity temperature dependence (Eq. 6) and using the rigorous intensity dependence derived from Eq. (S2) and (S5).

---

[3] Wenger, J.; Gérard, D.; Dintinger, J.; Mahboub, O.; Bonod, N.; Popov, E.; Ebbesen, T. W.; Rigneault, H. Emission and Excitation Contributions to Enhanced Single Molecule Fluorescence by Gold Nanometric Apertures. Opt. Express, 2008, 16, 3008–3020.